%% file: main.tex
\documentclass[conference]{IEEEtran}
\IEEEoverridecommandlockouts
\input{custom_style_iccad}
\begin{document}

\title{\ourtool: Fuzzing Processors with \\ Particle Swarm Optimization\\
}

\author{
{\rm Chen Chen$^{\ast,\dagger}$\thanks{$^\ast$These authors contributed equally to this work.}, Vasudev Gohil$^{\ast,\dagger}$, Rahul Kande$^\dagger$, Ahmad-Reza Sadeghi$^\ddagger$, and Jeyavijayan (JV) Rajendran$^\dagger$}\\ 
$^\dagger$Texas A\&M University, USA, $^\ddagger$Technische Universit\"at Darmstadt, Germany\\
{\tt $^\dagger$\{chenc, gohil.vasudev, rahulkande, jv.rajendran\}@tamu.edu,}\\
{\tt $^\ddagger$\{ahmad.sadeghi\}@trust.tu-darmstadt.de}
} 

\maketitle

\input{curr_version/abstract}


\input{curr_version/introduction}
\input{curr_version/background}
\input{curr_version/methodology}
\input{curr_version/experiment}
\input{curr_version/relatedwork}
\input{curr_version/discussion}
\input{curr_version/conclusion}

\bibliographystyle{IEEEtran}
\bibliography{IEEEabrv,Bibfile}

\end{document}

%% file: custom_style_iccad.tex
\usepackage{cite}
\usepackage{amsmath,amssymb,amsfonts}
\usepackage[ruled,vlined,linesnumbered]{algorithm2e}
\usepackage{graphicx}
\usepackage{diagbox}
\usepackage{multirow}
\usepackage{textcomp}
\usepackage{xcolor}
\usepackage{caption}
\usepackage{subcaption}
\usepackage{comment}
\usepackage{enumitem}

\usepackage{listings}
\input{listing_style}

\def\BibTeX{{\rm B\kern-.05em{\sc i\kern-.025em b}\kern-.08em
    T\kern-.1667em\lower.7ex\hbox{E}\kern-.125emX}}

\def\rahulen{1}
\if 1\rahulen

\else

\fi

\def\chencen{1}
\if 1\chencen

\else

\fi

\newcommand{\ourtool}{\textit{PSOFuzz}}
\newcommand{\mopt}{\textsc{MOpt}}

\newcommand{\thehuzz}{\textit{TheHuzz}}

\newcommand{\rc}{{\tt Rocket Core}}
\newcommand{\boom}{{\tt BOOM}}
\newcommand{\cva}{{\tt CVA6}}

\newcommand{\vcs}{\textit{VCS}}

\newcommand{\spike}{\textit{Spike}}
\newcommand{\chipyard}{\textit{Chipyard}}

\newcommand{\cvaSpdupVani}{$0.25$}
\newcommand{\cvaSpdupRst}{$6.39$}
\newcommand{\cvaSpdupSeed}{$2.22$}
\newcommand{\rcSpdupVani}{$0.10$}
\newcommand{\rcSpdupRst}{$0.23$}
\newcommand{\rcSpdupSeed}{$1.72$}
\newcommand{\boomSpdupVani}{$0.73$}
\newcommand{\boomSpdupRst}{$1.11$}
\newcommand{\boomSpdupSeed}{$1.25$}
\newcommand{\maxSpdupSeed}{$6.39$}

\newcommand{\cvaCovIncVani}{$-2.25$}
\newcommand{\cvaCovIncRst}{$2.00$}
\newcommand{\cvaCovIncSeed}{$1.00$}
\newcommand{\rcCovIncVani}{$-0.85$}
\newcommand{\rcCovIncRst}{$-0.06$}
\newcommand{\rcCovIncSeed}{$0.09$}
\newcommand{\boomCovIncVani}{$-0.15$}
\newcommand{\boomCovIncRst}{$0.01$}
\newcommand{\boomCovIncSeed}{$0.01$}
\newcommand{\maxCovInc}{$2$}

\newcommand{\maxVulDetSpd}{$15.25$}

%% file: listing_style.tex
\definecolor{codegreen}{rgb}{0,0.6,0}
\definecolor{codegray}{rgb}{0.5,0.5,0.5}
\definecolor{codepurple}{rgb}{0.58,0,0.82}
\definecolor{backcolour}{rgb}{0.95,0.95,0.92}

\let\othelstnumber=\thelstnumber
\def\createlinenumber#1#2{
    \edef\thelstnumber{%
        \unexpanded{%
            \ifnum#1=\value{lstnumber}\relax
              #2%
            \else}%
        \expandafter\unexpanded\expandafter{\thelstnumber\othelstnumber\fi}%
    }
    \ifx\othelstnumber=\relax\else
      \let\othelstnumber\relax
    \fi
}

\lstdefinestyle{customc}{
  belowcaptionskip=1\baselineskip,
  breaklines=true,
  frame=single,
  xleftmargin=0.35cm,
  xrightmargin=0.15cm,
  numbers=left,
  numbersep=5pt,  
  language=C,
  showstringspaces=false,
  basicstyle=\footnotesize\ttfamily,
  keywordstyle=\bfseries\color{green!40!black},
  commentstyle=\itshape\color{purple!40!black},
  identifierstyle=\color{blue},
  stringstyle=\color{orange},
}

\lstdefinestyle{customcArianeExploit1}{
  breaklines=true,
  frame=single,
  xleftmargin=0.4cm,
  xrightmargin=0.2cm,
  numbers=left,
  numbersep=5pt,  
  language=C,
  showstringspaces=false,
  basicstyle=\footnotesize\ttfamily,
  keywordstyle=\bfseries\color{green!40!black},
  commentstyle=\itshape\color{purple!60!black},
  identifierstyle=\color{blue},
  stringstyle=\color{yellow!50!black},
  morekeywords={asm},
  keywordstyle=[2]\bfseries\color{brown!60!black},
}

\lstdefinestyle{customcArianeExploit}{
  breaklines=true,
  frame=single,
  xleftmargin=0.4cm,
  xrightmargin=0.2cm,
  numbers=left,
  numbersep=5pt,  
  language=C,
  showstringspaces=false,
  basicstyle=\footnotesize\ttfamily,
  keywordstyle=\bfseries\color{blue},
  commentstyle=\itshape\color{green!50!black},
  identifierstyle=\color{black},
  stringstyle=\color{brown},
  morekeywords={asm},
  keywordstyle=[2]\bfseries\color{black},
}

\lstdefinestyle{customlog}{
  breaklines=true,
  frame=single,
  xleftmargin=0.35cm,
  xrightmargin=0.15cm,
  numbers=left,
  numbersep=5pt,  
  language=C,
  showstringspaces=false,
  basicstyle=\footnotesize\ttfamily,
  keywordstyle=\color{blue},
  commentstyle=\itshape\color{purple!40!black},
  identifierstyle=\color{blue},
  stringstyle=\color{orange},
  keywords=[2]{INFO},
  keywords=[3]{ERROR},x
  keywordstyle=[2]\bfseries\color{green!40!black},
  keywordstyle=[3]\bfseries\color{red!500!black},
}

\definecolor{verilogcommentcolor}{RGB}{0,124,0}
\definecolor{verilogkeywordcolor}{RGB}{49,49,255}
\definecolor{verilogsystemcolor}{RGB}{128,0,255}
\definecolor{verilognumbercolor}{RGB}{255,143,102}
\definecolor{verilogstringcolor}{RGB}{160,160,160}
\definecolor{verilogdefinecolor}{RGB}{128,64,0}
\definecolor{verilogoperatorcolor}{RGB}{0,0,128}
\definecolor{pointcolor}{RGB}{192,0,0} 
\lstdefinestyle{prettyverilog}{
   language           = Verilog,
   commentstyle       = \color{verilogcommentcolor},
   alsoletter         = \$'0123456789\`,
   literate           = *{+}{{\verilogColorOperator{+}}}{1}%
                         {-}{{\verilogColorOperator{-}}}{1}%
                         {@}{{\verilogColorOperator{@}}}{1}%
                         {;}{{\verilogColorOperator{;}}}{1}%
                         {*}{{\verilogColorOperator{*}}}{1}%
                         {?}{{\verilogColorOperator{? }}}{1}%
                         {:}{{\verilogColorOperator{:}}}{1}%
                         {<}{{\verilogColorOperator{<}}}{1}%
                         {>}{{\verilogColorOperator{>}}}{1}%
                         {!}{{\verilogColorOperator{!}}}{1}%
                         {^}{{\verilogColorOperator{^}}}{1}%
                         {|}{{\verilogColorOperator{| }}}{1}%
                         {||}{{\verilogColorOperator{|| }}}{1}%
                         {=}{{\verilogColorOperator{= }}}{1}%
                         {==}{{\verilogColorOperator{== }}}{1}%
                         {=>}{{\verilogColorOperator{=> }}}{1}%
                         {[}{{\verilogColorOperator{[}}}{1}%
                         {]}{{\verilogColorOperator{]}}}{1}%
                         {(}{{\verilogColorOperator{(}}}{1}%
                         {)}{{\verilogColorOperator{)}}}{1}%
                         {rightbracket}{{\verilogColorOperator{)}}}{1}%
                         {,}{{\verilogColorOperator{,}}}{1}%
                         {.}{{\verilogColorOperator{.}}}{1}%
                         {~}{{\verilogColorOperator{$\sim$}}}{1}%
                         {\%}{{\verilogColorOperator{\%}}}{1}%
                         {\&}{{\verilogColorOperator{\& }}}{1}%
                         {\&\&}{{\verilogColorOperator{\&\& }}}{1}%
                         {\#}{{\verilogColorOperator{\#}}}{1}%
                         {\ /\ }{{\verilogColorOperator{\ /\ }}}{3}%
                         {\ _}{\ \_}{2}%
                        ,
   morestring         = [s][\color{verilogstringcolor}]{"}{"},%
   identifierstyle    = \color{black},
   vlogdefinestyle    = \color{verilogdefinecolor},
   vlogconstantstyle  = \color{verilognumbercolor},
   vlogsystemstyle    = \color{verilogsystemcolor},
   basicstyle         = \small\fontencoding{T1}\ttfamily,
  columns=fullflexible, 
   keywordstyle       = \bfseries\color{verilogkeywordcolor},
   morekeywords      = {val, when, port, coverage, unique},
   numbers            = left,
   numbersep          = 5pt,
   tabsize            = 2,
   escapeinside       = {/*!}{!*/},
   upquote            = true,
   sensitive          = true,
   showstringspaces   = false, 
   frame              = single,
   breaklines         = true,
   abovecaptionskip   = 0pt,
   belowcaptionskip   = 2pt,   
   xleftmargin        =0.35cm,
   xrightmargin       =0.15cm,
   captionpos         = t,
   emph               = {Point, Point0, Point1, Point2, Point3, Point4, Point5, Point6, Point7, Point8, Point9},
   emphstyle          =\color{pointcolor},
   emph               = {[2] STVEC,SCOUNTEREN,MSTATUS,MTVEC,ML1_ICACHE_MISS,ML1_DCACHE_MISS,MITLB_MISS,MDTLB_MISS,
                             MLOAD,MSTORE,MEXCEPTION,MEXCEPTION_RET,MBRANCH_JUMP,MCALL,MRET,MMIS_PREDICT,MSB_FULL,
                             MIF_EMPTY,MHPM_COUNTER_17,MHPM_COUNTER_18,MHPM_COUNTER_19,MHPM_COUNTER_20,MHPM_COUNTER_21,
                             MHPM_COUNTER_22,MHPM_COUNTER_23,MHPM_COUNTER_24,MHPM_COUNTER_25,MHPM_COUNTER_26,MHPM_COUNTER_27,
                             MHPM_COUNTER_28,MHPM_COUNTER_29,MHPM_COUNTER_30,MHPM_COUNTER_31}, 
   emphstyle          = {[2]\bfseries\color{verilogkeywordcolor}}
}

\makeatletter

\newcommand\language@verilog{Verilog}
\expandafter\lst@NormedDef\expandafter\languageNormedDefd@verilog%
  \expandafter{\language@verilog}
  
\lst@SaveOutputDef{`'}\quotesngl@verilog
\lst@SaveOutputDef{``}\backtick@verilog
\lst@SaveOutputDef{`\$}\dollar@verilog

\newcommand\getfirstchar@verilog{}
\newcommand\getfirstchar@@verilog{}
\newcommand\firstchar@verilog{}
\def\getfirstchar@verilog#1{\getfirstchar@@verilog#1\relax}
\def\getfirstchar@@verilog#1#2\relax{\def\firstchar@verilog{#1}}

\newcommand\addedToOutput@verilog{}
\lst@AddToHook{Output}{\addedToOutput@verilog}

\newcommand\constantstyle@verilog{}
\lst@Key{vlogconstantstyle}\relax%
   {\def\constantstyle@verilog{#1}}

\newcommand\definestyle@verilog{}
\lst@Key{vlogdefinestyle}\relax%
   {\def\definestyle@verilog{#1}}

\newcommand\systemstyle@verilog{}
\lst@Key{vlogsystemstyle}\relax%
   {\def\systemstyle@verilog{#1}}

\newcount\currentchar@verilog
  
\newcommand\@ddedToOutput@verilog
{%
   \ifnum\lst@mode=\lst@Pmode%
      \expandafter\getfirstchar@verilog\expandafter{\the\lst@token}%
      \expandafter\ifx\firstchar@verilog\backtick@verilog
         \let\lst@thestyle\definestyle@verilog%
      \else
         \expandafter\ifx\firstchar@verilog\dollar@verilog
            \let\lst@thestyle\systemstyle@verilog%
         \else
            \expandafter\ifx\firstchar@verilog\quotesngl@verilog
               \let\lst@thestyle\constantstyle@verilog%
            \else
               \currentchar@verilog=48
               \loop
                  \expandafter\ifnum%
                  \expandafter`\firstchar@verilog=\currentchar@verilog%
                     \let\lst@thestyle\constantstyle@verilog%
                     \let\iterate\relax%
                  \fi
                  \advance\currentchar@verilog by \@ne%
                  \unless\ifnum\currentchar@verilog>57%
               \repeat%
            \fi
         \fi
      \fi
   \fi
}

\lst@AddToHook{PreInit}{%
  \ifx\lst@language\languageNormedDefd@verilog%
    \let\addedToOutput@verilog\@ddedToOutput@verilog%
  \fi
}

\newcommand{\verilogColorOperator}[1]
{%
  \ifnum\lst@mode=\lst@Pmode\relax%
   {\bfseries\textcolor{verilogoperatorcolor}{#1}}%
  \else
    #1%
  \fi
}

\makeatother

\lstdefinestyle{mystyle}{
    commentstyle=\textit,
    keywordstyle=\textbf,
    stringstyle=\color{codepurple},
    basicstyle=\ttfamily,
    breakatwhitespace=false,         
    breaklines=true,      
    frame=single, 
    framexleftmargin=\parindent,
    captionpos=b,                    
    keepspaces=true,                 
    numbers=left,    
    numberstyle=\normalsize,
    stepnumber=1,
    numbersep=5pt,   
    xleftmargin=1.5\parindent,
    showspaces=false,                
    showstringspaces=false,
    showtabs=false,                  
    tabsize=2
}

%% file: curr_version/abstract.tex
\begin{abstract}
     Hardware security vulnerabilities in computing systems compromise the security defenses of not only the hardware but also the software running on it. Recent research has shown that hardware fuzzing is a promising technique to efficiently detect such vulnerabilities in large-scale designs such as modern processors. However, the current fuzzing techniques do not adjust their strategies dynamically toward faster and higher design space exploration, resulting in slow vulnerability detection, evident through their low design coverage. 

     To address this problem, we propose~\ourtool, which uses particle swarm optimization (PSO) to schedule the mutation operators and to generate initial input programs dynamically with the objective of detecting vulnerabilities quickly. Unlike traditional PSO, which finds a single optimal solution, we use a modified PSO that dynamically computes the optimal solution for selecting mutation operators required to explore new design regions in hardware. We also address the challenge of inefficient initial seed generation by employing PSO-based seed generation.
     Including these optimizations, our final formulation outperforms fuzzers without PSO. Experiments show that \ourtool{} achieves up to $\mathbf{15.25 \times}$ speedup for vulnerability detection and up to $\mathbf{2.22 \times}$ speedup for coverage compared to the state-of-the-art simulation-based hardware fuzzer.
     
\end{abstract}

%% file: curr_version/introduction.tex
\section{Introduction}~\label{sec:intro}
Hardware designs are becoming increasingly complex to meet the rising need for custom hardware and increased performance. However, as the design complexity grows, verification dominates the development lifecycle---over 70\% of resources are spent
to verify the security of a hardware design~\cite{fine2003coverage}. Therefore, a verification technique that can expose design flaws/vulnerabilities as early as possible is critical~\cite{rajendran2015detecting,dessouky2019hardfails,sadeghi2021organizing,chen2022trusting}.

Recent research has shown that coverage feedback-based fuzzers can effectively detect security vulnerabilities in modern processors and achieve coverage faster than traditional hardware verification techniques such as random regression\cite{rfuzz,muduli2020hyperfuzzing,canakci2021directfuzz,hur2021difuzzrtl,fuzzhwlikesw,kande2022thehuzz,ragab_bugsbunny_2022,chen2023hypfuzz}.
Fuzzers start with seed inputs and apply predefined mutation operators on current inputs to generate new inputs.
Fuzzers utilize one or several coverage metrics to track the activity caused by the inputs in the hardware and guide themselves to generate inputs that explore new hardware regions, thereby accelerating vulnerability detection. These fuzzers successfully found vulnerabilities that execute undefined behaviors, compromise memory isolation, lead to privilege escalations, etc., in hardware designs.
Due to their promising potential, semiconductor giants such as Intel and Google are actively developing new and efficient hardware fuzzers for vulnerability detection~\cite{presifuzz,fuzzhwlikesw}.

Though hardware fuzzing is promising, its coverage increment usually stagnates quickly, leaving design spaces unexplored and vulnerabilities undetected. One of the main reasons is that all the existing hardware fuzzers schedule the mutation operators and generate seeds using static schemes, and lack feedback-guided dynamic updates via interaction with the coverage achieved and the complexity of the hardware (more details in Sec.~\ref{sec:rel})~\cite{rfuzz, hur2021difuzzrtl,kande2022thehuzz,fuzzhwlikesw,ragab_bugsbunny_2022,canakci2021directfuzz,muduli2020hyperfuzzing,chen2023hypfuzz}. For example, \cite{hur2021difuzzrtl} generates input instructions with the same operands to verify the renaming strategies in out-of-order processors. However, this strategy is not optimal for processors without renaming strategies. While \cite{kande2022thehuzz} uses profiling to analyze the correlation between the mutation operators and the processor instructions, the scheduling is static throughout fuzzing and does not update with coverage achieved. 
Therefore, to dynamically adjust the schedule of mutation operators and seed generation, we develop an approach to equip any hardware fuzzer with particle swarm optimization~(PSO) targeting faster vulnerability detection and coverage achievement.

PSO is a bio-inspired algorithm that uses a swarm consisting of multiple particles to find the optimal solution in a high-dimensional space~\cite{pso}. 
Since fuzzers usually run multiple threads in parallel to accelerate the design space exploration, we model the scheduler of each thread's mutation operations as a particle. The probability distribution of each mutation operator being selected composes the position of a particle. 
Each particle updates its position based on the highest coverage achieved by itself and the highest coverage achieved by the swarm in the current iteration.

Researchers have used PSO in software fuzzing to find the optimal probability distribution of mutation operators~\cite{lyu2019mopt}. Results show a significant improvement over the fuzzers without PSO. However, one cannot simply use their technique for hardware fuzzing for the following reasons: (i) They assume a static optimal probability distribution of the mutation operators, which is not true for hardware fuzzing. Our experiments show that if a static distribution is used, two critical vulnerabilities are not detected even after using PSO (more details in Sec.~\ref{sec:solution1} and~\ref{sec:exp}). (ii) They use randomly generated programs as seeds, which is not ideal for hardware fuzzing (more details in Sec.~\ref{sec:solution2}).
To successfully apply PSO to hardware fuzzing, we must address two critical challenges: 
\begin{itemize}[leftmargin=*]
    \item \textbf{Challenge 1: Saturation of Particles' Performance.} Naively applying PSO to hardware fuzzing results in the particles converging to a fixed position leading to a lack of coverage improvement and vulnerabilities undetected.
    \item \textbf{Challenge 2: Ineffective Seed Generation.} The seeds from which the fuzzing loop starts strongly affect the coverage achieved. Traditional fuzzers use either statically or randomly generated seeds, which results in slower coverage increments. Moreover, the performance of applying PSO to hardware fuzzing is also subject to the seed: poor quality seeds result in poor coverage.
\end{itemize}
We address these challenges by designing specific solutions (in Sec.~\ref{sec:solution1} and \ref{sec:solution2}) that result in faster vulnerability detection and coverage achievement. In particular, we augment PSO with a reset strategy to schedule mutation operators dynamically. Moreover, we employ PSO to select high-quality seed inputs resulting in faster coverage.
Overall, the main contributions of this work are as follows:
\begin{enumerate}[leftmargin=*]
    \item To the best of our knowledge, we develop the first technique that uses the PSO algorithm to schedule the mutation operators and generate seed inputs in hardware fuzzing.
    \item We overcome challenges in adapting PSO to hardware fuzzers. In particular, we develop optimizations to reset the particles for selecting mutation operators and a novel PSO-based seed generation algorithm for hardware fuzzers.
    \item We evaluate \ourtool{} on three widely-used open-source RISC-V processors and achieve up to \maxVulDetSpd $\times$ speedup in detecting vulnerabilities and up to $2.22\times$ speedup in achieving coverage compared to the state-of-the-art simulation-based fuzzer without PSO.
\end{enumerate}

%% file: curr_version/background.tex
\section{Background}

\subsection{Hardware Processor Fuzzers}~\label{sec:fuzz_bg}
Hardware processor fuzzers iteratively generate testing inputs (i.e., \textbf{inputs} or \textbf{tests}) such as binary executables to detect vulnerabilities in target hardware (i.e., design-under-test~(DUT)). These fuzzers mainly consist of a \textit{seed generator}, \textit{mutation engine}, \textit{feedback engine}, and \textit{vulnerability detector}~\cite{hur2021difuzzrtl, kande2022thehuzz, chen2023hypfuzz, ragab_bugsbunny_2022}.
The \textit{seed generator} generates an initial set of tests called \textbf{seeds}. Fuzzer simulates the hardware with these seeds to generate feedback and output.
The \textit{feedback engine} captures the activity in hardware, such as toggling of the value of flip-flops~\cite{kande2022thehuzz}, as \textbf{coverage} data during simulation.
The \textit{mutation engine} modifies the tests that achieve new coverage (i.e., \textbf{interesting} tests) to create new tests. For example, the \textit{Bitflip} mutation operator flips one random bit in the test~\cite{citeafl}. The \textit{vulnerability detector} detects vulnerabilities in hardware by comparing its output with the output of a \textbf{golden-reference model}~(GRM)~\cite{kande2022thehuzz,hur2021difuzzrtl,chen2023hypfuzz}. 

Existing hardware fuzzers use static mutation operator scheduling and seed generation schemes which cannot dynamically adjust based on the complexity of hardware and the coverage achieved, slowing down vulnerability detection and design space exploration (more details in Sec.~\ref{sec:method} and \ref{sec:rel}).

\subsection{PSO Algorithm}
PSO is a bio-inspired algorithm that allocates multiple particles to iteratively search for an optimal solution~\cite{pso}. The position of a particle is a candidate solution and is updated based on its velocity. The velocity is determined by the best position the particle ever achieved and the best position the entire swarm ever achieved. A particle's velocity, $v_i$, and position, $p_i$, are updated as
\begin{align}\label{eq:up_v}
\begin{split}
    v_i(t+1) &= k\times v_i(t) + r_1\times(l_i^{best}(t)-p_i(t))\\
             &+ r_2 \times (g^{best}(t) - p_i(t))
\end{split}
\end{align}
\begin{equation}\label{eq:up_p}
    p_i(t+1) = p_i(t) + v_i(t+1) \tag{2}
\end{equation}

Here, $l_i^{best}(t)$ represents the best position the particle $i$ ever achieved since the beginning (i.e., $t = 0$), and $g^{best}(t)$ represents the best position ever achieved by any particle in the swarm. $k$ is a pre-defined constant, and $r_1$ and $r_2$ are two random displacement weights that decide how fast the particle $i$ will move toward $l_i^{best}(t)$ and $g^{best}(t)$, respectively. 

PSO has been widely applied to optimize the performance of applications due to its searching efficiency and convergence speed~\cite{zhan2009adaptive}. 
Though the solutions found by PSO are heuristic results, they are usually close to the real global optimum~\cite{shi1998modified}.
The software community has also leveraged PSO~(e.g., \mopt{}~\cite{lyu2019mopt}) to detect vulnerabilities (e.g., memory crashes) in programs. 
However, for hardware fuzzing, the optimal probability distribution of the mutation operators changes with coverage achieved so far, and the quality of seeds greatly affects the performance of the fuzzer. Hence,~\mopt{} cannot be applied for hardware fuzzing directly.

%% file: curr_version/methodology.tex
\section{\ourtool{}: Fuzzing Processors with PSO}\label{sec:method}
In this section, we first formulate the problem of selecting optimal mutation operators as a PSO problem. However, the formulation has critical limitations, such as saturation of particles' performance and inefficient seed generation. We analyze these limitations and devise appropriate solutions, resulting in a final formulation that outperforms fuzzers without PSO.

\subsection{Selecting Mutation Operators Using PSO}\label{sec:PSO_mutation_prelim_formulation}
Most existing hardware fuzzers select mutation operators uniformly at random, i.e., they have an equal probability of selecting from all possible mutation operators~\cite{rfuzz,kande2022thehuzz,chen2023hypfuzz, canakci2021directfuzz, muduli2020hyperfuzzing, ragab_bugsbunny_2022,fuzzhwlikesw, hur2021difuzzrtl}. 
This is not ideal as mutation operators have varying impacts on covering new points in hardware. The following example demonstrates this.

\begin{lstlisting}[language=Verilog, label = {listing:example_listing}, caption={Code snippet from \cva{}~\cite{cva6} processor (simplified to improve readability)},style=prettyverilog,float,belowskip=-15pt,aboveskip=0pt,firstnumber=1,linewidth=\linewidth]
unique case (instr.opcoderightbracket
  ...
    CSRRS: begin
      if (instr.rs1 == 5'b0rightbracket
        instruction_o.op = CSR_READ;// c1
      else
        instruction_o.op = CSR_SET; // c2
\end{lstlisting}
\textbf{Motivational Example.} Consider a component from the decoder module in the \cva{}~\cite{cva6} processor as shown in Listing~\ref{listing:example_listing}. 
When decoding the \texttt{CSRRS rd, csrReg, rs1} instruction, the processor performs \texttt{CSR\_READ} (line 5) or \texttt{CSR\_SET} (line 7) operation on the control and status register~(CSR), \texttt{csrReg}, based on the register operand \texttt{rs1} (line 4).
Assume the coverage points $c_1$ and $c_2$ in the \textit{if-else} block as shown in line 5 and line 7.
Suppose the instruction in the current input is \texttt{CSRRS x16, stvec, x0}. 
The value of \texttt{rs1} for this instruction is \texttt{0}. Hence the \textit{if} condition evaluates to \texttt{True}, and the \texttt{CSR\_READ} operation is performed on the CSR, \texttt{stvec}, covering the point $c_1$. 
Now, our objective, when mutating this input, is to cover $c_2$. Also, suppose we have only two mutation operators in this toy example: \textit{Bitflip}, which flips one bit of the operand, \texttt{rs1}, and \textit{OpcodeMut}, which mutates the opcode to another opcode. If both mutation operators are equally likely to be selected, the probability of covering $c_2$ after mutation is $(0.5\times 0) + (0.5\times 1) = 0.5$. On the other hand, if we assign a higher weight to \textit{Bitflip}, say $0.9$, the probability of covering $c_2$ is $(0.1\times 0) + (0.9\times 1) = 0.9$. So, selecting mutation operators with equal probability is not ideal.
To address this problem, we formulate the problem of finding the optimal weights for the mutation operators as a PSO problem, as explained next.

\textbf{Notation.} Let $M$ be an ordered list of all mutation operators (i.e., $M[j]$ is the $j^{\text{th}}$ mutation operator).
Let $W^M = [w^M_1, w^M_2, \ldots, w^M_{|M|}]$ be a weight vector for the $|M|$ operators. So, $w^M_j$ is the weight of the $j^{\text{th}}$ mutation operator $M[j]$. Note that since we interpret the weights as the probabilities of selecting the mutation operators, we normalize the weights so that $\sum_{j=1}^{|M|} w^M_j = 1$ and $w^M \geq 0, \forall w^M \in W^M$.

\textbf{Particles.} We associate a particle with one thread of the fuzzer. Since fuzzers run multiple, say $|N|$, threads simultaneously, we have a swarm of $|N|$ particles associated with the fuzzer. Each of these $|N|$ threads has a different weight vector $W_i^M$ for mutating the inputs in its thread. Each such $W_i^M$ is assigned as the position $p_i^M$ of the corresponding particle $i^M$ ($M$ indicates that the particle refers to mutation operators). Mathematically, $W_i^M = p_i^M$.

\textbf{Local Best Position.} Following PSO, we assign the local best position, $l_i^{best,M}$, of a particle as the best position 
of that particle so far. In our case, we define a position at iteration $t_2$, $p_i^M(t_2)$, as better than the position at iteration $t_1$, $p_i^M(t_1)$, if and only if $p_i^M(t_2)$ results in higher coverage than $p_i^M(t_1)$.

\textbf{Global Best Position.} Likewise, following PSO, we assign a single global best position, $g^{best,M}$, as the best position of all particles in the swarm.

\begin{figure}[t]
    \centering
    \includegraphics[width=0.85\columnwidth,trim=18 18 18 19, clip]{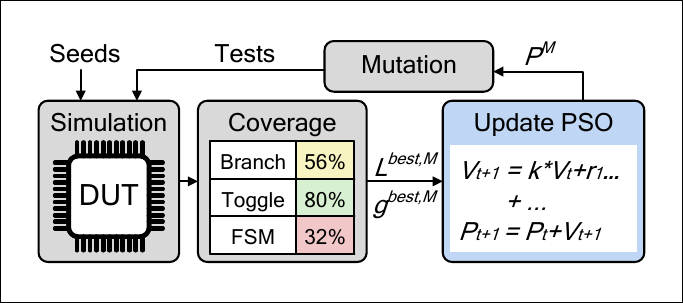}
    \caption{Flow for integrating PSO in TheHuzz.}
    \label{fig:integrating_PSO_in_TheHuzz}
\end{figure}

Fig.~\ref{fig:integrating_PSO_in_TheHuzz} illustrates the high-level flow of this formulation. Like other hardware fuzzers, we generate a set of seeds, which are simulated to obtain their coverage information. Then, we use this coverage information to update the $L^{best,M} = [l_1^{best,M}, l_2^{best,M}, \ldots, l_{|N|}^{best,M}]$ and the $g^{best,M}$. Then, we update the velocities ($v_i^M$) and positions ($p_i^M$) of all particles according to Eqs.~(\ref{eq:up_v}) and~(\ref{eq:up_p}). 
We sample the mutation operators according to the updated positions ($p_i^M$), i.e., mutation operator weights ($W_i^M$), and perform the mutations to generate new tests. These new tests are simulated to obtain coverage, and the cycle continues. After several iterations, the particles are expected to converge to the optimal solution.

This preliminary solution provides a way to find a weight distribution of the mutation operators. However, it has two critical limitations, as explained below.

\begin{figure*}[t]
    \centering
    \includegraphics[width=\textwidth,trim={0 0.3cm 0 0.3cm},clip]{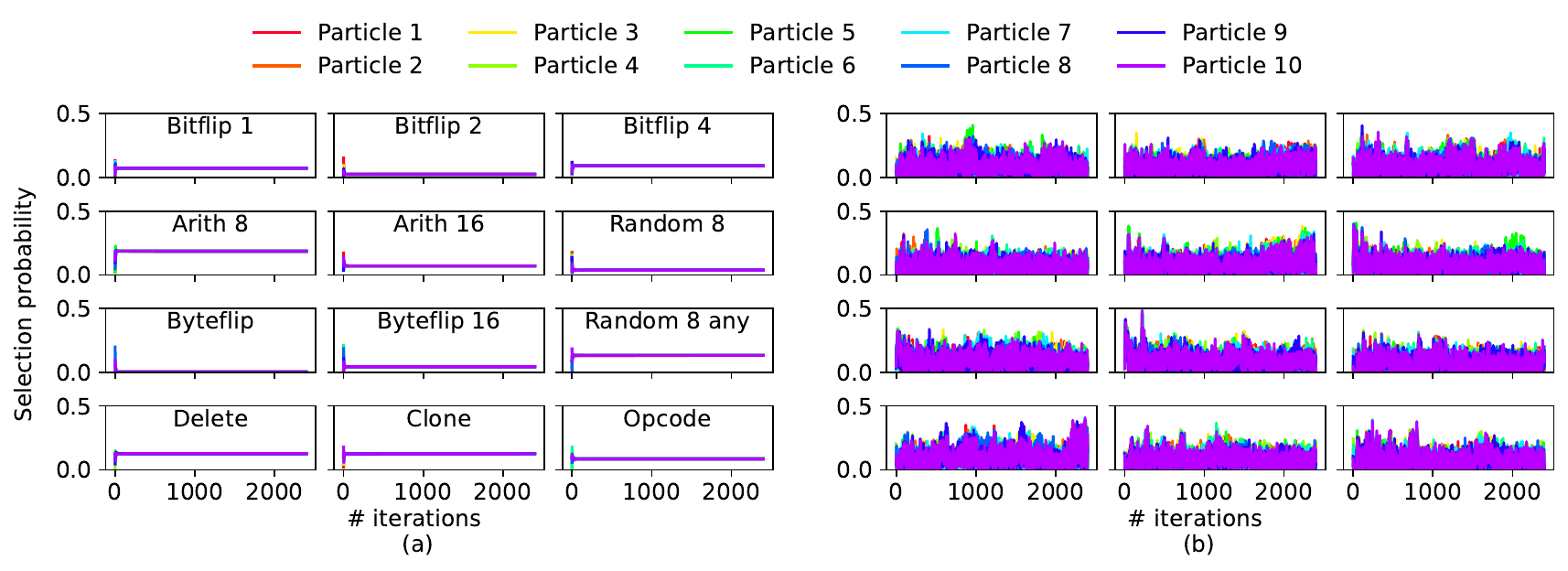}
    \caption{Selection probability trend for all mutation operators for \texttt{CVA6}~\cite{cva6} (a) without and (b) with reset}
    \label{fig:cva6_particle_position_103_evolution_without_and_with_reset}
\end{figure*}
\subsection{Reset Particles and Seeds}\label{sec:solution1}

\textbf{Challenge 1: Saturation of Particles' Performance.} Traditional PSO is designed so that the particles converge to an optimum (ideally the global optimum) in the solution space. However, for fuzzing hardware, one must find new coverage points in each iteration that have not been covered. In other words, the optimal solution (i.e., the weights of the mutation operators) changes based on the coverage achieved so far, as explained in the following example.

\textbf{Motivational Example.} 
Refer to Listing.~\ref{listing:example_listing} with two coverage points, $c_1$ and $c_2$. Without loss of generality, assume two mutation operators $m_1$ and $m_2$ with the following property: $\mathcal{P}(c_1|m_1)=0.9$ and $\mathcal{P}(c_1|m_2) = 0.2$, where $\mathcal{P}(c_i|m_j)$ denotes the probability of covering $c_i$ conditioned on the event that mutation operator $m_j$ is used.\footnote{Note that since the coverage points belong to two different mutually exclusive and exhaustive branches, $\mathcal{P}(c_2|m_1)=0.1$ and $\mathcal{P}(c_2|m_2) = 0.8$.} Suppose during the first iteration ($t=1$), the weights for the two mutation operators are equal, i.e., $0.5$. Then, the probability of covering $c_1$, $\mathcal{P}(c_1)^{t=1} = (0.5 \times 0.9) + (0.5 \times 0.2) = 0.55$, and $\mathcal{P}(c_2)^{t=1} = (0.5 \times 0.1) + (0.5 \times 0.8) = 0.45$. Now, if $c_1$ is covered in the first iteration, then, for the second iteration ($t=2$), having equal weights for both operators results in $\mathcal{P}(c_2)^{t=2} = \mathcal{P}(c_2)^{t=1} = 0.45$. On the other hand, if we have a larger weight for $m_2$, say $0.9$, then $\mathcal{P}(c_2)^{t=2} = (0.1 \times 0.1) + (0.9 \times 0.8) = 0.73$. In fact, to maximize the likelihood of covering $c_2$ in the second iteration, the weight for $m_2$ should be $1$. 

This simple example shows that to maximize the likelihood of covering the maximal number of points with the minimal number of iterations (i.e., input tests), the weights of the mutation operators should be decided dynamically based on the coverage achieved so far.
However, during our experiments with the preliminary PSO formulation, we observe that the positions of the particles ($p_i^M$) 
saturate after a few iterations because their velocities ($v_i^M$) become zero according to Eq.~(\ref{eq:up_v}). In other words, the weights of selecting the mutation operators ($W_i^M$) stagnate very quickly.
Fig.~\ref{fig:cva6_particle_position_103_evolution_without_and_with_reset}~(a) demonstrates this phenomenon for the \texttt{CVA6} processor. The particles saturate within $50$ iterations. Due to this, the preliminary formulation fails to detect critical vulnerabilities (more details in Sec.~\ref{sec:exp}).

\textbf{Solution 1.} To address challenge 1, we reset the position ($p_i^M$) and velocity ($v_i^M$) of the particles that saturate, i.e., that do not yield new coverage. Resetting removes saturating particles and replaces them with new particles, leading to new weights for the mutation operators ($W_i^M$). Along with resetting the position and velocity of the saturated particle, we also reset the associated seed because the position and velocity of a particle are a function of the seed it started from (as seen in Fig.~\ref{fig:integrating_PSO_in_TheHuzz}). Using the same seed when resetting the particle will likely lead to the exploration of design space that has already been explored by that particle. This is suboptimal since our objective is to cover new regions in the design.

The condition for resetting the particles and their seeds is decided based on the coverage trend in recent iterations. We reset when there is no improvement in coverage for $\beta^M$ consecutive iterations. Here, $\beta^M$ controls the trade-off between runtime and exploitation of learned knowledge by the particles. Larger values of $\beta^M$ can lead to a deeper exploration of design but will incur runtime overhead since the particles are not reset quickly after they saturate.
On the other hand, smaller values of $\beta^M$ can reduce runtime but will make the algorithm similar to random exploration, which will not cover deeper regions of the design. 

Fig.~\ref{fig:cva6_particle_position_103_evolution_without_and_with_reset}~(b) shows the positions of the particles for different mutation operators after implementing this solution for \texttt{CVA6}. When the particles saturate, they are reset (along with the associated seeds), resulting in higher coverage speedup (as seen in Table~\ref{tab:cov}). 
Algorithm~\ref{alg:reset} details how this solution is incorporated using a reset monitor, \textit{RstMon}. It takes as input a set of particles, their $L^{best}$, resetting threshold ($\beta$), a variable to count the number of consecutive iterations with no improvement ($Ct$), and a function $\mathcal{F}$ that returns the fitness (i.e., coverage) of a particle. 
The algorithm iterates over all particles and compares the coverage with the particle's local best coverage (lines 2 and 3). If a particle's coverage is more than its local best coverage, the local best is updated, and the counter ($Ct$) is reset (lines 4 and 5). Otherwise, the counter is incremented, indicating no improvement in coverage for one more consecutive iteration for that particle (line 7). If the counter exceeds the threshold $\beta$ for a particle, that particle is added to the set of particles to be reset, $rst\_P$ (lines 8 and 9). Finally, the algorithm updates the global best ($g^{best}$) using all particles' local bests (line 10).

\input{Codes/pseudoalgo/alg1}

\begin{figure*}[t]
    \centering
    \includegraphics[trim=18 21 18 19,clip,width=0.9\textwidth]{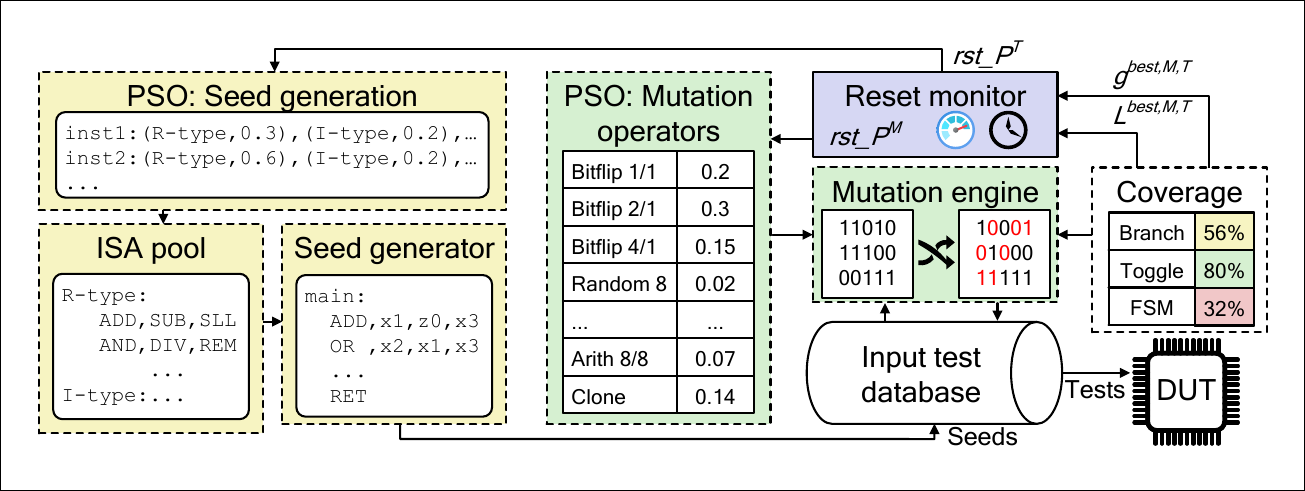}
    \caption{\ourtool~framework.}
    \label{fig:PSOHuzz_framework}
\end{figure*}
\subsection{Seed Generation Using PSO}\label{sec:solution2}

\textbf{Challenge 2: Ineffective Seed Generation.} 
A drawback of existing hardware fuzzers (as well as our preliminary formulation) is that they use a static probability distribution of instructions to generate the seeds~\cite{rfuzz,kande2022thehuzz,chen2023hypfuzz, canakci2021directfuzz, muduli2020hyperfuzzing, ragab_bugsbunny_2022}. 
However, the optimal set of instructions required by the seeds to trigger new coverage points changes with time during fuzzing.\footnote{An example similar to the one described in Sec.~\ref{sec:solution1} can demonstrate this, but we omit it in the interest of space.} 
Moreover, the performance of each particle for mutation operators ($i^M$), i.e., the number of iterations it survives, is highly related to the quality of its seed. 

\textbf{Solution 2.} To address this problem of static distribution of instructions for seed generation, we devise a dynamic seed generation algorithm using PSO. The idea is to map the problem of identifying optimal probabilities of the instructions for seeds as a PSO problem. Next, we discuss this formulation.

\textbf{Notation.} Let $T$ be an ordered list of instruction types, such as R-type and I-type~\cite{riscv_home}. Let $W^T = [w^T_1, w^T_2, \ldots, w^T_{|T|}]$ be the weight vector for the $|T|$ instruction types. So, $w^T_j$ is the weight of the $j^{\text{th}}$ instruction type $T[j]$, and $\sum_{j=1}^{|T|} w^T_j = 1$ and $w^T \geq 0, \forall w^T \in W^T$. 

\textbf{Particles.} Similar to the PSO formulation for selecting mutation operators (Sec.~\ref{sec:PSO_mutation_prelim_formulation}), we associate a particle with one thread of the fuzzer. So, there are a total of $|N|$ particles in the swarm, where $|N|$ is the number of fuzzer threads. Suppose that each seed consists of a sequence of $|O|$ instructions.
Then, for each of  the $|N|$ threads being run simultaneously, we have a different weight vector $W_i^T$ 
(which contains $|O|\times |T|$ entries, one for each of the $|T|$ instruction types for each of the $|O|$ instructions in the seed).

Each weight vector $W_i^T$ is assigned as the position $p_i^T$ of the corresponding particle $i^T$. Here, the $T$ in the superscript indicates that the particle is for the instruction types, as opposed to the $M$ in the particles for selecting mutation operators in Sec.~\ref{sec:PSO_mutation_prelim_formulation}.
Mathematically, $W_i^T = p_i^T$.

\textbf{Local Best Position.} Following PSO, we assign the local best position, $l_i^{best,T}$, of a particle as the best position (i.e., the probabilities of the instruction types) of that particle so far. Since this PSO is for seed generation, the objective it aims to maximize is the quality of the generated seed. 
We measure the quality of a particle for selecting instruction types ($i^T$) as the number of iterations the associated particles for selecting mutation operators ($i^M$) survive.
For instance, for a given particle, $i^T$, if at iteration $t_1$ and  $t_2$, its associated particle for selecting mutation operators ($i^M$) survived for $150$ and $200$ iterations, respectively (as explained in Sec.~\ref{sec:solution1}), then $p_{i}^T(t_2)$ is better than $p_{i}^T(t_1)$.

\textbf{Global Best Position.} Likewise, following PSO, we assign a single global best position, $g^{best,T}$, as the best position of all particles in the swarm.

The high-level flow of this PSO for seed generation is very similar to that of the PSO for selecting mutation operators. We start with randomly initialized particles and update their velocities and positions according to Eqs.~(\ref{eq:up_v}) and (\ref{eq:up_p}), respectively. Moreover, learning from the prior pitfall of saturation of particles' performance (Sec.~\ref{sec:solution1}), we also reset the particles for this PSO-based seed generation when they saturate. 
Since the quality of a particle for selecting instruction types ($i^T$) is measured by the number of iterations its corresponding particle for selecting mutation operators ($i^M$) survives, we reset $i_T$ when its corresponding $i^M$ is reset $\beta^T$ consecutive times.
Similar to $\beta^M$, $\beta^T$ controls the trade-off between runtime and exploitation of learned knowledge by $i^T$s.

\subsection{Putting it All Together}

Fig.~\ref{fig:PSOHuzz_framework} shows the \ourtool{} framework, which includes PSO formulation along with the resetting of seeds and particles for selecting mutation operators as well as the PSO formulation for seed generation.
The PSO formulation of seed generation guides the \textit{seed generator} by selecting the instruction type from the instruction set architecture (ISA) pool. 
Once the instruction type is sampled, the \textit{seed generator} randomly selects an opcode of that instruction type and then the operands to create the instruction. These instructions are then used to create the seeds, which are added to the \textit{input test database}. 
The DUT is simulated with the tests to obtain coverage. 
This coverage data is used to update the particles' velocities and positions through their local and global bests and the reset monitor. The updated positions are used to generate seeds and mutate tests in the subsequent iterations.

\input{Codes/pseudoalgo/alg2}
Algorithm~\ref{alg:overall} details the steps of~\ourtool. It takes the DUT, a user-defined time limit, $t_{limit}$, a user-defined target coverage, $tc$, the thresholds for resetting the PSOs for mutation operators and seed generation, $\beta^M$ and $\beta^T$, and the constant, $k$, used to update the velocities of the particles (Eq.~(\ref{eq:up_v})). We fuzz the DUT and return the coverage achieved, $cov$. 
First, we initialize all required variables for PSO ($P^M,V^M,P^T,V^T,L^{best,M},g^{best,M},L^{best,T},g^{best,T}$). We also initialize 
the saturation counters ($Ct^M, Ct^T$) along with the sets indicating particles that need to be reset because of saturation ($rst\_P^M,rst\_P^T$). 
Next, on line 7, we use the positions $P^T$ to generate the initial set of seed $tests$.
Then, the main fuzzing loop starts on line 8, which iterates until either the target coverage ($tc$) is achieved or the runtime exceeds the time limit, $t_{limit}$.
In each iteration of the fuzzing loop, we first simulate the DUT with the generated $tests$ to obtain the function $\mathcal{F}^M$ (which calculates the fitness of the particles)
and the coverage achieved so far, $cov$ (line 9). 
Next, on line 10, this $\mathcal{F}^M$ is used to calculate the local and global bests for the particles for mutation operators and to decide which particles need to be reset according to $\beta^M$ following Algorithm~\ref{alg:reset}.
Then, skipping ahead on line 15, the velocities and the positions of the particles for mutation operators are updated, 
and on lines 16 and 17, the new $tests$ are generated for both particles that are reset
and not reset, respectively.

Additionally, on line 11, we check if any of the particles for the mutation operators needs to be reset, i.e., $rst\_P^M \neq \phi$, and
calculate the number of iterations those particles survived (i.e., fitness) and return $\mathcal{F}^T$ on line 12.
This $\mathcal{F}^T$ is used to calculate the local and global bests for the particles for seed generation and to decide which ones to reset according to $\beta^T$ on line 13.
Finally, on line 14, the velocities and the positions of the particles for seed generation are updated.

%% file: Codes/pseudoalgo/alg1.tex
\begin{algorithm}[t]
\SetFuncSty{textsc}
\SetKwFunction{GenSeed}{GenSeed}
\SetKwProg{Fn}{Function}{:}{}
\DontPrintSemicolon
\caption{\textit{RstMon}: Reset monitor}\label{alg:reset}
\KwIn{
    $P, L^{best}, \beta, Ct, \mathcal{F}$
}
\KwOut{
    $L^{best},g^{best},Ct, rst\_P$
}

$rst\_P \gets \phi$\;

\For{$i \in P$}{
    \If{$\mathcal{F}(p_i) > \mathcal{F}(L^{best}[i])$}{
        $L^{best}[i] \gets p_i$\;
        $Ct[i] \gets 0$\;
    }\Else{
        $Ct[i] \gets Ct[i] + 1$\;
    }
    \tcp{add particle to the reset set}
    \If{$Ct[i] > \beta$} {
        $rst\_P \gets rst\_P \cup \{i\}$\;
    }
}

$g^{best} \gets \text{particle with highest } L^{best}$

return $L^{best},g^{best},rst\_P,Ct$\;

\end{algorithm}

%% file: Codes/pseudoalgo/alg2.tex
\begin{algorithm}[t]
\SetFuncSty{textsc}
\SetKwFunction{GenSeed}{GenSeed}
\SetKwProg{Fn}{Function}{:}{}
\DontPrintSemicolon
\caption{\ourtool{}}\label{alg:overall}
\KwIn{
    $DUT,t_{limit},tc,\beta^M,\beta^T,k$
}
\KwOut{$cov$: total coverage achieved}
\tcp{initialization}
$t \gets 0$, $cov \gets 0$, $tests \gets \phi$\; 
Initialize $P^M, V^M, P^T, V^T$\;
Initialize $L^{best,M}$ to $P^M$ and $g^{best,M}$ to $p_0^M$\;
Initialize $L^{best,T}$ to $P^T$ and $g^{best,T}$ to $p_0^T$\;


$Ct^M \gets 0$, $Ct^T \gets 0$\;
$rst\_P^M \gets \phi$, $rst\_P^T \gets \phi$\;
$tests \gets GenSeed(P^T, tests, rst\_P^M)$\;
\While(\tcp*[f]{main loop}){$(cov < tc)$ and $(t < t_{limit})$} {
    $\mathcal{F}^M,  cov \gets Simulate(DUT, tests)$\;
    $L^{best,}, g^{best,M}, rst\_P^M, Ct^M \gets RstMon(P^M, L^{best,M}, \beta^M, Ct^M, \mathcal{F}^M)$\;
    \If(\tcp*[f]{update PSO seed}){$rst\_P^M \neq \phi$}{
        $\mathcal{F}^T \gets CalcFitness(rst\_P^M, P^T)$\;
    
        $L^{best,T}, g^{best,T}, rst\_P^T, Ct^T \gets RstMon(P^T, L^{best,T}, \beta^T, Ct^T, \mathcal{F}^T)$\;        
        
        $P^T, V^T \gets UpdatePV(P^T, V^T, L^{best,T}, g^{best,T}, rst\_P^T, k)$
    }
    $P^M, V^M\gets UpdatePV(P^M, V^M, L^{best,M}, g^{best,M}, rst\_P^M, k)$
    \tcp*[f]{gen. seeds for reset particles}\;
    $tests \gets GenSeed(P^T, tests, rst\_P^M)$\;
    \tcp{mutate tests for other particles}
    $tests \gets Mut(P^M, tests, \{i^M|i^M \notin rst\_P^M\})$\;
}
return $cov$\;

\end{algorithm}

%% file: curr_version/experiment.tex
\section{Evaluation}\label{sec:exp}

\input{tables/table1}

In this section, we evaluate the speed of~\ourtool{} in detecting vulnerabilities and achieving coverage on three popular open-sourced RISC-V~\cite{riscv_home} processors.

\subsection{Evaluation Setup}
We use the state-of-the-art simulation-based processor fuzzer, \thehuzz{}~\cite{kande2022thehuzz} as a baseline for all evaluations.\footnote{Following~\cite{kande2022thehuzz}, we implemented~\thehuzz~in Python.} 
To ensure that the enhancements in results exclusively stem from PSO adaptation, we integrate PSO schemes into \thehuzz{}'s \textit{mutation engine} and \textit{seed generator}.
In addition to reporting the results for our final formulation,~\ourtool, which is also denoted as \textbf{PSO+Reset+Seed}, we also report the results for (i) the preliminary formulation, \textbf{PSO}, which includes the PSO-based mutation operator selection, but neither the resetting solution (Solution 1), nor the PSO-based seed generation technique (Solution 2), and (ii) the intermediate formulation, \textbf{PSO+Reset}, which includes the PSO-based mutation operator selection with reset, but not PSO-based seed generation.

Following~\cite{kande2022thehuzz}, we set the number of simultaneous threads and hence the number of particles, $|N|$, as $10$ and the number of instructions in each program, $|O|$, as $20$. To have a balanced impact of exploration (due to velocity) and exploitation (due to local and global bests), we set $k$ (Eq.~(\ref{eq:up_v})) as $0.5$. Based on empirical observations, we set $\beta^M = \beta^T = 3$, as a larger $\beta$ will cause runtime overhead, while a smaller $\beta$ will make \ourtool{} generate test cases similar to random exploration, as discussed in Sec.~\ref{sec:solution1}.

Since most commercial processors are close-sourced, we evaluate \ourtool{} using three widely-used open-sourced processors from RISC-V~\cite{riscv_home} ISA: \cva{}~\cite{cva6}, \boom{}~\cite{boom}, and \rc{}~\cite{rocket_chip_generator}. Existing hardware processor fuzzers use similar open-sourced processors for their evaluation~\cite{hur2021difuzzrtl,kande2022thehuzz,chen2023hypfuzz}. \cva{} supports out-of-order execution~(OoO) and a custom single instruction-multiple data~(SIMD) floating point unit. 
\rc{} is an in-order processor. 
\boom{} is a superscalar processor built using the components of \rc{}. 
Moreover, all of these processors are capable of booting Linux operating system while \cva{} is commercially used to build processors for high-security applications~\cite{migv}.
Therefore, these processors with different complexity form a diverse benchmark set to evaluate \ourtool{}. 

We use Synopsys \vcs{}~\cite{vcs} as the simulation tool and \chipyard{}~\cite{chipyard} as the system-on-chip~(SoC) simulation environment. To ensure a fair comparison, we ran experiments on each benchmark for 24 hours and repeated them thrice. 
We demonstrate our results using the branch coverage metric, which is highly related to vulnerability detection~\cite{mockus2009test}. We use a Linux-based CPU running at 2.6GHz, with 64 threads and 512GB of RAM for our experiments.

\subsection{Vulnerability Detection}
The vulnerability detection strategy used by our techniques (PSO, PSO+Reset, PSO+Reset+Seed) compares the processor's outputs with that of a GRM.
Many existing fuzzers use a similar detection strategy~\cite{kande2022thehuzz,hur2021difuzzrtl,chen2023hypfuzz,ragab_bugsbunny_2022}. 
The strategy compares the architectural states of the processor and the GRM for each executed instruction.
The architectural states include the instructions committed; any exceptions triggered; the general purpose registers~(GPRs) modified; the CSRs modified; the privilege level of the processor; and the memory address and data accesses. A mismatch in architectural states indicates a potential vulnerability in the processor. Similar to \cite{kande2022thehuzz,hur2021difuzzrtl,chen2023hypfuzz,ragab_bugsbunny_2022}, we use \spike{}~\cite{spike}, the RISC-V ISA emulator, as the GRM.

\textbf{Comparison with \thehuzz{}.}
Table~\ref{tb:b_list_v1} shows the number of tests and the speedup of our techniques compared to \thehuzz. 
\ourtool{} detects all the vulnerabilities detected by \thehuzz{} up to \maxVulDetSpd $\times$ faster.
The results also show the necessity of using the reset strategy for PSO. Na\"ive PSO is slower than \thehuzz{} in detecting vulnerabilities because its particles are likely exploring the design spaces that have already been explored, as mentioned in Sec.~\ref{sec:solution1}. 
PSO+Reset performs better than PSO due to resetting the particles, allowing it to dynamically select the optimal mutation operators. It also outperforms PSO+Reset+Seed (\ourtool{}) in detecting some vulnerabilities, such as V3, because \ourtool{} initially spends time exploring the optimal probabilities for seed generation. 
However, \ourtool{} achieves the highest speedup compared to \thehuzz{} and other PSO techniques for most vulnerabilities because it uses the optimal probabilities for both seed generation and instruction mutation, efficiently exploring different regions of the design space. 

\input{tables/table2}

\begin{figure*}[htb!]
    \centering
    \includegraphics[width=\textwidth]{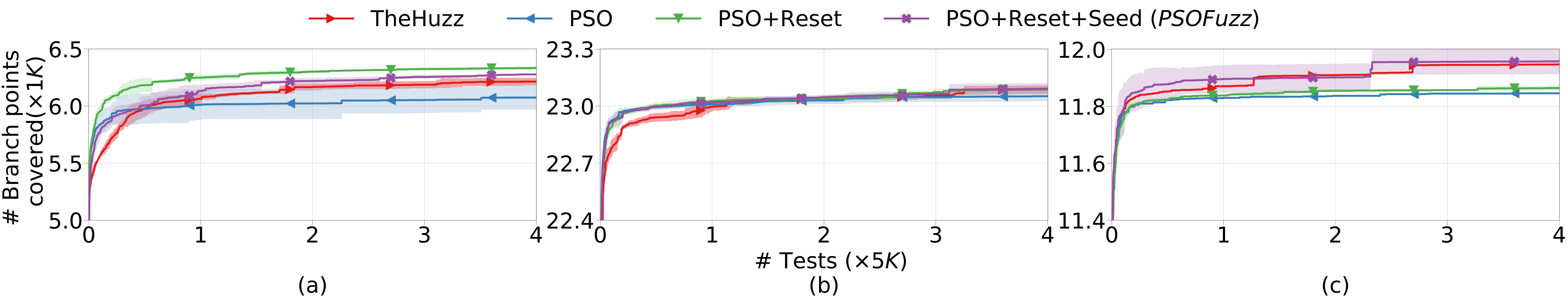}
    \caption{Coverage achieved compared to \thehuzz{}~\cite{kande2022thehuzz} for (a) \cva{}~\cite{cva6}, (b) \boom{}~\cite{boom}, and (c) \rc{}~\cite{rocket_chip_generator}.}
    \label{fig:tot_branch}
\end{figure*}

\subsection{Coverage Achievement}
We now compare the capability of our techniques
in achieving coverage against \thehuzz{}, as coverage achieved is a proxy metric to the extent of design verified and determines its readiness to tape out~\cite{verifiwhitepaper}. Across the three processors, PSO optimizations achieve up to \maxSpdupSeed{}$\times$ speedup compared to \thehuzz{} while obtaining up to \maxCovInc\% more coverage after fuzzing for 24 hours with more than 20K tests, as shown in Table~\ref{tab:cov}. Fig.~\ref{fig:tot_branch} shows the mean and standard deviation of coverage achieved with respect to the number of input tests.

\ourtool{} outperforms PSO+Reset on \rc{} and \boom{} because the PSO-based seed generation dynamically adjusts the optimal set of instructions required to cover new points. 
While \ourtool{} is slower than PSO+Reset on \cva{}, it still outperforms \thehuzz{} and PSO techniques. 

Among the three processors, \ourtool{} achieves the fastest speedup (\cvaSpdupSeed{}$\times$) and highest coverage increment (\cvaCovIncSeed{}\%) on \cva{}, which is the complex processor with OoO and SMID features. 
This is because the PSO optimizations aid in covering the hard-to-reach coverage points in the custom floating point unit of \cva{}, which are difficult to cover with static mutation operator scheduling schemes.
Even on the remaining processors, \boom{} and \rc{}, \ourtool{} achieves faster coverage (up to $\rcSpdupSeed{}\times$) compared to \thehuzz{} and covers more points than \thehuzz{}. The speedup on these processors is less than that of \cva{} because \rc{} is a simple processor compared to \cva{} while \boom{} is built by duplicating most of the components of \rc{}. Hence, \thehuzz{} covers more than $87\%$ of their points, resulting in less scope for improvement for \ourtool{}.

In summary, \ourtool{} detects vulnerabilities and achieves coverage faster than \thehuzz{} on all three processors, demonstrating the necessity of dynamic scheduling of mutation operators and seed generation. 

%% file: tables/table1.tex
\begin{table*}[!hbt]
\caption{Vulnerability detection speedup compared to \thehuzz{}~\cite{kande2022thehuzz}. N.D. denotes ``not detected.''}
\label{tb:b_list_v1}
\resizebox{\textwidth}{!}{%
\begin{tabular}{|c|c|c|c|cc|cc|cc|}
\hline
\multirow{2}{*}{Processor} & \multirow{2}{*}{Vulnerability} & \multirow{2}{*}{CWE} & \thehuzz{}~\cite{kande2022thehuzz} & \multicolumn{2}{c|}{PSO} & \multicolumn{2}{c|}{PSO+Reset} & \multicolumn{2}{c|}{PSO+Reset+Seed (\ourtool)} \\ \cline{4-10} 
 &  &  & \# Tests & \multicolumn{1}{c|}{\# Tests} & Speedup & \multicolumn{1}{c|}{\# Tests} & Speedup & \multicolumn{1}{c|}{\# Tests} & Speedup \\ \hline
\multirow{7}{*}{\cva{}~\cite{cva6}} & \begin{tabular}[c]{@{}c@{}}V1: Access \textit{invalid} addresses\\ without throwing exceptions.\end{tabular} & CWE-1252 & $3.52\times10^2$ & \multicolumn{1}{c|}{$1.38\times10^2$} & $2.55\times$ & \multicolumn{1}{c|}{$5.71\times10^2$} & $0.62\times$ & \multicolumn{1}{c|}{$2.32\times10^2$} & $1.52\times$ \\ \cline{2-10} 
 & \begin{tabular}[c]{@{}c@{}}V2: Decode multiplication\\ instructions incorrectly.\end{tabular} & CWE-440 & $2.71\times10^3$ & \multicolumn{1}{c|}{$5.09\times10^3$} & $0.53\times$ & \multicolumn{1}{c|}{$1.80\times10^3$} & $1.50\times$ & \multicolumn{1}{c|}{$4.51\times10^2$} & $6.01\times$ \\ \cline{2-10} 
 & \begin{tabular}[c]{@{}c@{}}V3: Access unimplemented \\ CSRs and return X-values.\end{tabular} & CWE-1281 & $4.75\times10^3$ & \multicolumn{1}{c|}{N.D.} & N.D. & \multicolumn{1}{c|}{$8.56\times10^2$} & $5.54\times$ & \multicolumn{1}{c|}{$5.70\times10^3$} & $0.83\times$ \\ \cline{2-10} 
 & \begin{tabular}[c]{@{}c@{}}V4: Cache coherency\\ violation undetected.\end{tabular} & CWE-1202 & $1.83\times10^2$ & \multicolumn{1}{c|}{$1.23\times10^4$} & $0.01\times$ & \multicolumn{1}{c|}{$1.70\times10^1$} & $10.76\times$ & \multicolumn{1}{c|}{$1.20\times10^1$} & $\mathbf{15.25\times}$ \\ \cline{2-10} 
 & \begin{tabular}[c]{@{}c@{}}V5: Decode the \texttt{FENCE.I}\\ instruction incorrectly.\end{tabular} & CWE-440 & $6.80\times10^{1}$ & \multicolumn{1}{c|}{$4.50\times10^2$} & $0.15\times$ & \multicolumn{1}{c|}{$1.80\times10^1$} & $3.78\times$ & \multicolumn{1}{c|}{$1.35\times10^2$} & $0.50\times$ \\ \cline{2-10} 
 & \begin{tabular}[c]{@{}c@{}}V6: Incorrect exception type\\ in instruction queue.\end{tabular} & CWE-1202 & $5.55\times10^3$ & \multicolumn{1}{c|}{$1.94\times10^4$} & $0.29\times$ & \multicolumn{1}{c|}{$5.39\times10^2$} & $10.30\times$ & \multicolumn{1}{c|}{$1.26\times10^3$} & $4.40\times$ \\ \cline{2-10} 
 & \begin{tabular}[c]{@{}c@{}}V7: Some \textit{illegal} instructions\\ without throwing exceptions.\end{tabular} & CWE-1242 & $9.30\times10^1$ & \multicolumn{1}{c|}{N.D.} & N.D. & \multicolumn{1}{c|}{$2.54\times10^2$} & $0.37\times$ & \multicolumn{1}{c|}{$7.40\times10^1$} & $1.26\times$ \\ \hline
\rc{}~\cite{rocket_chip_generator} & \begin{tabular}[c]{@{}c@{}}V8: \texttt{EBREAK} does not\\ increase instruction count.\end{tabular} & CWE-1201 & $1.67\times10^3$ & \multicolumn{1}{c|}{$6.67\times10^3$} & $0.25\times$ & \multicolumn{1}{c|}{$1.07\times10^3$} & $1.56\times$ & \multicolumn{1}{c|}{$4.10\times10^2$} & $4.07\times$ \\ \hline 
\end{tabular}%
}
\end{table*}

%% file: tables/table2.tex
\begin{table*}[t]
\caption{Coverage improvement compared to \thehuzz{}~\cite{kande2022thehuzz}.}
\resizebox{\textwidth}{!}{%
\begin{tabular}{|c|c|ccc|ccc|ccc|}
\hline
\multirow{2}{*}{Core} & \thehuzz{}~\cite{kande2022thehuzz} & \multicolumn{3}{c|}{PSO} & \multicolumn{3}{c|}{PSO+Reset} & \multicolumn{3}{c|}{PSO+Reset+Seed (\ourtool)} \\ \cline{2-11} 
 & Total & \multicolumn{1}{l|}{Total} & \multicolumn{1}{l|}{Increment} & Speedup & \multicolumn{1}{c|}{Total} & \multicolumn{1}{c|}{Increment} & \multicolumn{1}{l|}{Speedup} & \multicolumn{1}{c|}{Total} & \multicolumn{1}{c|}{Increment} & \multicolumn{1}{l|}{Speedup} \\ \hline
\cva{}~\cite{cva6} & $6214$ & \multicolumn{1}{c|}{$6074$} & \multicolumn{1}{c|}{\cvaCovIncVani $\%$} & \cvaSpdupVani $\times$ & \multicolumn{1}{c|}{$6334$} & \multicolumn{1}{c|}{\cvaCovIncRst $\%$} & \cvaSpdupRst $\times$ & \multicolumn{1}{c|}{$6277$} & \multicolumn{1}{c|}{\cvaCovIncSeed $\%$} & $\mathbf{2.22}\times$ \\ \hline
\boom{}~\cite{boom} & $23086$ & \multicolumn{1}{c|}{$23052$} & \multicolumn{1}{c|}{\boomCovIncVani $\%$} & \boomSpdupVani $\times$ & \multicolumn{1}{c|}{$23088$} & \multicolumn{1}{c|}{\boomCovIncRst $\%$} & \boomSpdupRst $\times$ & \multicolumn{1}{c|}{$23092$} & \multicolumn{1}{c|}{\boomCovIncSeed $\%$} & \boomSpdupSeed $\times$ \\ \hline
\rc{}~\cite{rocket_chip_generator} & $11947$ & \multicolumn{1}{c|}{$11846$} & \multicolumn{1}{c|}{\rcCovIncVani $\%$} & \multicolumn{1}{c|}{\rcSpdupVani $\times$} & \multicolumn{1}{c|}{$11864$} & \multicolumn{1}{c|}{\rcCovIncRst $\%$} & \rcSpdupRst $\times$ & \multicolumn{1}{c|}{$11958$} & \multicolumn{1}{c|}{\rcCovIncSeed $\%$} & \rcSpdupSeed $\times$ \\ \hline
\end{tabular}%
}
\label{tab:cov}
\end{table*}


%% file: curr_version/relatedwork.tex
\section{Related Work}\label{sec:rel}
This section describes the existing hardware fuzzers, their limitations, and how \ourtool{} addresses those limitations. 

\textbf{\textit{RFUZZ}}~\cite{rfuzz} uses mux-toggle coverage to capture the activity in the select signals of MUXes. It schedules mutation operators statically, similar to AFL fuzzer~\cite{citeafl}. Also, mux-toggle coverage is not scalable to large designs such as \boom{}~\cite{hur2021difuzzrtl}. \textbf{\textit{HyperFuzzing}}~\cite{muduli2020hyperfuzzing} is an SoC fuzzer that defines security properties and uses AFL fuzzer's static mutation operator scheduling and random seed generation to detect vulnerabilities. \textbf{\textit{DirectFuzz}}~\cite{canakci2021directfuzz} modifies \textit{RFUZZ} to perform directed fuzzing on specific modules of the target hardware. However, it uses similar static schemes as \textit{RFUZZ}. 

Processor fuzzer \textbf{\textit{DIFUZZRTL}}~\cite{hur2021difuzzrtl} uses activity in the control registers as coverage to address the scalability issues of \textit{RFUZZ}. 
It uses grammar-based mutation operators inspired by AFL. But, it generates seeds randomly and uses static mutation operator scheduling schemes. 
\textbf{\textit{BugsBunny}}~\cite{ragab_bugsbunny_2022} fuzzes specific target signals in the hardware using \textit{DIFUZZRTL} as the base fuzzer. Hence, its mutation operator scheduling and seed generation schemes are static. 
\textbf{\textit{Trippel et al.}}~\cite{fuzzhwlikesw} convert hardware to a software model and fuzz using AFL fuzzer with static mutation scheduling. 
\textbf{\textit{TheHuzz}}~\cite{kande2022thehuzz} is another processor fuzzer that uses different types of code coverage metrics to capture activity in the combinational and sequential logic of the hardware. 
It also uses a profiler to optimize the mutation scheduling. However, this optimization is static for a given processor and does not change with the coverage achieved. 

A recent hardware fuzzer \textbf{\textit{HyPFuzz}}~\cite{chen2023hypfuzz} attempts to address the slower coverage speeds of hardware fuzzers by combing formal tools with fuzzing. This optimization to fuzzing is orthogonal to \ourtool{} as \textit{HyPFuzz} also uses static mutation operator scheduling and seed generation schemes which can be optimized through our PSO techniques. 

In summary, all of the existing hardware fuzzers 
use static mutation operator scheduling and seed generation schemes and cannot dynamically adjust their strategies for the target hardware and coverage achieved. In contrast, we propose \ourtool, a PSO-guided fuzzer, to demonstrate how hardware fuzzers can be equipped with PSO optimizations to improve their vulnerability detection and coverage achievement speeds.

%% file: curr_version/discussion.tex
\section{Discussion}\label{sec:discu}
In this section, we discuss possible extensions to improve \ourtool~and hardware fuzzing in general.

\textbf{Dynamically Identifying Optimal Solutions.} 
Though the reset strategy works well in practice, it is a heuristic that modifies the PSO algorithm. 
In the future, we plan to model and analyze the impact of resetting threshold ($\beta$) on the capability of vulnerability detection and design space exploration of \ourtool{}. Moreover, we will also evaluate the performance of other meta-heuristic algorithms, such as simulated annealing~\cite{bertsimas1993simulated} and differential evolution~\cite{price2013differential}, and identify the ideal one for hardware fuzzing.

\textbf{Leveraging Prior Knowledge for Initialization.} 
One key way to improve its performance is by initializing with a good starting point~\cite{simpson2017penalising}. \ourtool{} initializes particles' positions randomly. However, we can use the static schemes of existing fuzzers to initialize the particles. 
For example, the static probabilities from the profiling state of \thehuzz{} can be used as the initial position of particles to select the mutation operators. 

%% file: curr_version/conclusion.tex
\section{Conclusion}\label{sec:concl}
Processor fuzzers, which rely on the mutation of input tests, have shown promise in detecting vulnerabilities. However, they use static scheduling strategies for mutation and seed generation, which is not ideal. To overcome this limitation, we develop a novel PSO-based framework,~\ourtool{}, that can be integrated with any fuzzer to find optimal scheduling strategies. To ensure better vulnerability detection performance, we augment na\"ive PSO with (i) a reset strategy that allows it to find dynamic selection probabilities for mutation operators and (ii) a PSO-based seed generation algorithm. Experimental results with the state-of-the-art simulation-based hardware fuzzer demonstrate that~\ourtool~can speed up vulnerability detection by up to $15.25\times$ and coverage achievement by up to $2.22\times$. In conclusion,~\ourtool~addresses a critical limitation of existing processor fuzzers and significantly improves the effectiveness and efficiency of processor verification.

\section{Acknowledgement}
Our research work was partially funded by the US Office of Naval Research (ONR Award \#N00014-22-1-2279), by Intel's Scalable Assurance Program, and by the European Union (ERC, HYDRANOS, 101055025).
Any opinions, findings, conclusions, or recommendations expressed herein are those of the authors and do not necessarily reflect those of the US Government, the European Union, or the European Research Council. 